\documentclass[sigconf,natbib=true,anonymous=false]{acmart}

\usepackage{pifont} 
\usepackage{xcolor}
\usepackage{color}
\usepackage{graphicx}
\usepackage{float}
\usepackage{subfigure}
\usepackage{threeparttable}
\usepackage{bm}

\usepackage{ulem}

\usepackage{multirow}

\usepackage{enumitem}
\usepackage{ulem}
\usepackage{rotating}

\theoremstyle{definition}

\usepackage{amsmath}

\usepackage{graphicx}
\usepackage{extarrows}
\usepackage{enumitem}

\AtBeginDocument{%
  \providecommand\BibTeX{{%
    \normalfont B\kern-0.5em{\scshape i\kern-0.25em b}\kern-0.8em\TeX}}}

\copyrightyear{2024}
\acmYear{2024}
\setcopyright{acmlicensed}
\acmConference[SIGIR '24] {Proceedings of the 47th International ACM SIGIR Conference on Research and Development in Information Retrieval}{July 14-18, 2024}{Washington D.C., USA}
\acmBooktitle{Proceedings of the 47th International ACM SIGIR Conference on Research and Development in Information Retrieval (SIGIR '24), July 14-18, 2024, Washington D.C., USA}
\acmPrice{15.00}
\acmISBN{978-1-4503-9408-6/23/07}
\acmDOI{10.1145/XXXXXX.XXXXXX}

\begin{document}

\title{NFARec: A Negative Feedback-Aware Recommender Model}
\author{Xinfeng Wang}
\email{g22dtsa7@yamanashi.ac.jp}
\affiliation{%
\institution{University of Yamanashi}
\orcid{0000-0003-4491-8369}
\city{Kofu}
\country{Japan}
}

\author{Fumiyo Fukumoto}
\email{fukumoto@yamanashi.ac.jp}
\affiliation{%
  \institution{University of Yamanashi}
  \orcid{0000-0001-7858-6206}
\city{Kofu}
\country{Japan}
}

\author{Jin Cui}
\email{g22dtsa5@yamanashi.ac.jp}
\affiliation{%
  \institution{University of Yamanashi}
  \orcid{0000-0001-9575-3678}
\city{Kofu}
\country{Japan}
}

\author{Yoshimi Suzuki}
\email{ysuzuki@yamanashi.ac.jp}
\affiliation{%
  \institution{University of Yamanashi}
  \orcid{0000-0001-5466-7351}
\city{Kofu}
\country{Japan}
}

\author{Dongjin Yu}
\email{yudj@hdu.edu.cn}
\affiliation{%
  \institution{Hangzhou Dianzi University}
  \orcid{0000-0001-8919-1613}
\city{Hangzhou}
\country{China}
}

\copyrightyear{2024}
\acmYear{2024}
\setcopyright{acmlicensed}\acmConference[SIGIR '24]{Proceedings of the 47th International ACM SIGIR Conference on Research and Development in Information Retrieval}{July 14--18, 2024}{Washington, DC, USA}
\acmBooktitle{Proceedings of the 47th International ACM SIGIR Conference on Research and Development in Information Retrieval (SIGIR '24), July 14--18, 2024, Washington, DC, USA}
\acmDOI{10.1145/3626772.3657809}
\acmISBN{979-8-4007-0431-4/24/07}
\acmPrice{}
\renewcommand{\shortauthors}{Xinfeng Wang et al.}

\begin{abstract}

Graph neural network (GNN)-based models have been extensively studied for recommendations, as they can extract high-order collaborative signals accurately which is required for high-quality recommender systems. However, they neglect the valuable information gained through negative feedback in two aspects: (1) different users might hold opposite feedback on the same item, which hampers optimal information propagation in GNNs, and (2) even when an item vastly deviates from users' preferences, they might still choose it and provide a negative rating. In this paper, we propose a negative feedback-aware recommender model (NFARec) that maximizes the leverage of negative feedback. To transfer information to multi-hop neighbors along an optimal path effectively, NFARec adopts a feedback-aware correlation that guides hypergraph convolutions (HGCs) to learn users' structural representations. Moreover, NFARec incorporates an auxiliary task - predicting the feedback sentiment polarity (i.e., positive or negative) of the next interaction - based on the Transformer Hawkes Process. The task is beneficial for understanding users by learning the sentiment expressed in their previous sequential feedback patterns and predicting future interactions. Extensive experiments demonstrate that NFARec outperforms competitive baselines. Our source code and data are released at \href{https://github.com/WangXFng/NFARec}{https://github.com/WangXFng/NFARec}.

\begin{figure}[t]
  \includegraphics[width=\linewidth]{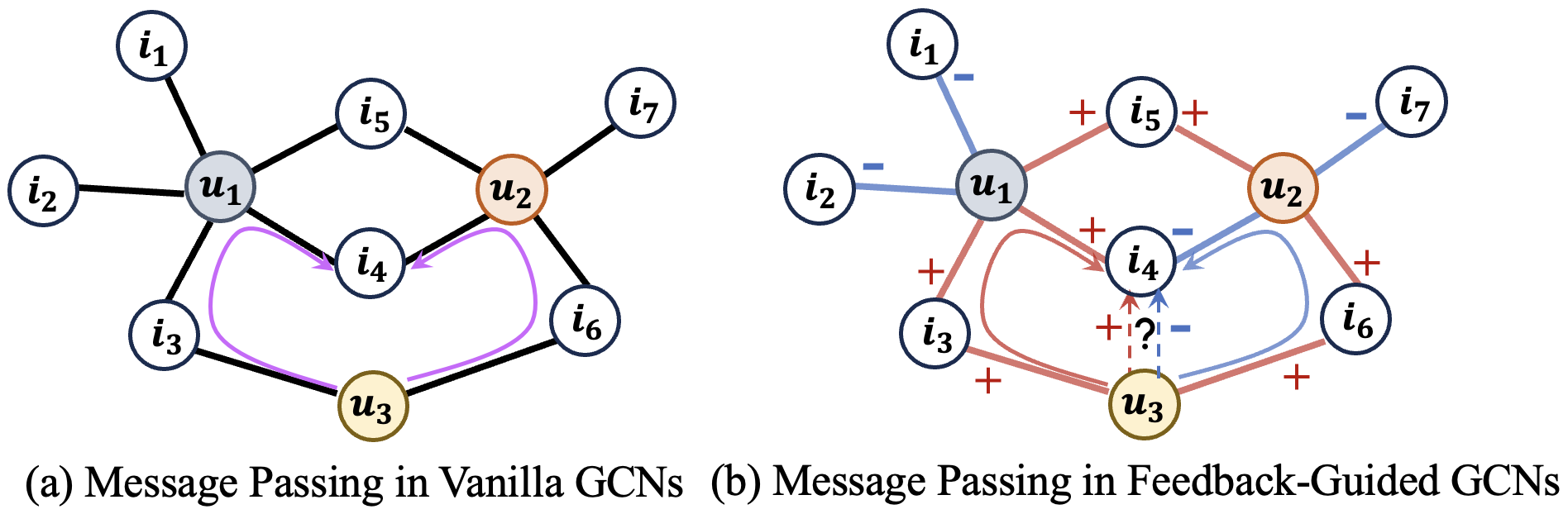}
  \caption{Illustration of motivation. In (a), vanilla GCN operators transmit messages from $u_3$ to $i_4$ through their interactive connections. However, the optimal message-passing path, either ($u_3$ $\xrightarrow{}$ $i_3$ $\xrightarrow{}$ $u_1$ $\xrightarrow{}$ $i_4$) or ($u_3$ $\xrightarrow{}$ $i_6$ $\xrightarrow{}$ $u_2$ $\xrightarrow{}$ $i_4$), is determined by the user, $u_1$ or $u_2$, with greater similarity to $u_3$ in (b). }
  \label{fig:motivation}
\end{figure}

\end{abstract}



\maketitle

\section{Introduction}

With the advent of the internet era, effective recommender systems bring considerable benefits to both service supporters and customers in various fields, such as social service, movie, and book recommendations \cite{wu2021self,  yu2022graph, wang2022towards, yu2022low}.
Graph convolution networks (GCNs) have been widely employed to capture user\textendash item interactive signals for recommendations due to their effectiveness in capturing high-order structural features \cite{jin2020multi, he2020lightgcn, hao2021pre}. Several authors have incorporated self-supervised learning \cite{yu2021socially, yu2021self}, hypergraph correlations \cite{yu2021self, xia2022hypergraph, xia2022self}, and contrastive augmentations \cite{cai2022lightgcl, lin2022improving, yu2023xsimgcl} into neural graph learning to boost the representation capability of node embeddings. 
However, they do not pay attention to the fact that individual users may give different or conflicting feedback for the same item, making GCNs generate flawed node representations. For instance, in Fig. \ref{fig:motivation}, both user $u_1$ and $u_2$ exhibit similarity with user $u_3$. This is because $u_1$ and $u_3$ provide positive feedback on item $i_3$, and both $u_2$ and $u_3$ give positive feedback on item $i_6$. However, there is a divergence in their preferences when it comes to item $i_4$; $u_1$ provides positive feedback, while $u_2$ provides negative feedback. 
Typically, recommender algorithms assess whether user $u_1$ or $u_2$ is more similar to $u_3$ to predict the preferences of user $u_3$ for item $i_4$.
If $u_3$ expresses a positive sentiment score toward item $i_4$, the path ($u_3$ $\xrightarrow{}$ $i_3$ $\xrightarrow{}$ $u_1$ $\xrightarrow{}$ $i_4$) should be the optimal message-passing than the path ($u_3$ $\xrightarrow{}$ $i_6$ $\xrightarrow{}$ $u_2$ $\xrightarrow{}$ $i_4$). The observations show that when treating neighboring nodes equally in GCNs, conducting convolution along the connections between nodes results in a lack of distinction among nodes. This suggests that user feedback is an important factor to guide message-passing in GCNs. 
%

Negative feedback from interactive events, such as low ratings \cite{huang2023negative, wang2021interactive}, clicking dislikes \cite{zhao2018recommendations, wu2020neural, gong2022positive}, or skipping content \cite{huang2023negative, gong2022positive}, have been utilized to generate negative representations for recommendations. 
%
%
However, their approaches learn the negative and positive representations independently. This results in their methods failing to effectively capture behavioral features across users' interacted items. Even though these items receive diverse positive and negative feedback, there still exists some connection among them that leads to interactions with the same user. 
Several recent studies either exclude negative instances, such as ratings below 4 out of 5 \cite{yu2021socially, yu2020enhancing, yu2022graph}, or regard all interaction records as positive feedback. These approaches either simplify the task, which deviates from real-world scenarios, or overlook crucial feedback features to enrich user representations, leading to a performance decrease.
In contrast, our solution is to utilize interaction feedback to guide GCN operators to learn both behavioral features and user feedback.

User sequential patterns, such as time-aware patterns \cite{chen2022intent}, long- and short-term user behaviors \cite{wang2021sequential, chang2021sequential}, and spatiotemporal relationships \cite{luo2021stan, sun2024maran}, are extensively employed to improve sequential recommendations \cite{tao2022sminet, jin2020multi, yang2022getnext, xie2022contrastive, wang2023sequential}.
%
More recently, several efforts \cite{xia2022self, xia2022hypergraph, wang2023eedn} have been made to integrate sequential and graph-based representations for recommendations. Despite their inspiring results, they overlook the fact that even though an item significantly diverges from a user's preferences, the user may still try it and provide negative feedback. The issue prevents the model from understanding user behavioral characteristics. For instance, some customers prefer products with lower prices over reputations, while others may prefer to be cautious and only engage with highly rated items after a thorough evaluation. Hence, it is beneficial for understanding users by learning the sentiment expressed in their previous sequential patterns of feedback, together with 
predicting whether they will interact with an item or not.

Motivated by the aforementioned issues, we propose a negative feedback-aware recommender model (NFARec) that maximizes the use of negative feedback in sequential and structural patterns. For sequential patterns, the NFARec incorporates an auxiliary task - predicting the feedback sentiment polarity of the next interaction - based on the Transformer Hawkes Process (THP). This is motivated by the advantage of the THP to model the influence of past events on the current event, whose impact diminishes over time.
%
As such, the task prompts the NFARec to recognize the items that the user interacted with but may not prefer and learn the emotional relatedness between those interacted items to understand users.

To enhance the modeling of structural patterns, we propose a two-phase hypergraph convolution (HGC) approach that leverages high-order feedback relations between users and items. In the initial stage, we employ HGCs to capture correlations beyond pairwise interactions among users and items via a user hypergraph, in which a hyperedge denotes a user and nodes in the hyperedge refer to the items interacted by the user. Then, we guide the HGC operator to efficiently exchange messages between neighboring nodes during convolutions through a high-order feedback-aware path.
%
The main contributions of our work can be summarized as follows:

\begin{itemize}

\item We propose a negative feedback-aware recommender model that maximizes the leverage of negative feedback in sequential and structural patterns.

\item We adopt an auxiliary task to learn the sentiment relatedness between the user's interacted items based on THP. We empirically find that the task promotes the NFARec to identify items the user has interacted with but may not prefer.

\item We propose a two-phase HGC approach that leverages feedback relations between users and items to construct an optimal path for message-passing in HGC.

\item Extensive experiments on five public real-world datasets demonstrate that our NFARec model outperforms SOTA recommender methods.

\end{itemize}

\section{Related Work}

\noindent
\textbf{Sequential Neural Network for Recommendation.}
Early efforts \cite{kang2018self, li2020time} concentrate on modeling user sequential trajectories for recommendations. For instance, \citet{wang2021sequential} propose to mine periodic and short-term patterns from sequential behaviors. \citet{luo2021stan} exploit spatiotemporal information of the historical check-ins. \citet{yang2022getnext} present a user-agnostic global trajectory flow map to learn discontinuous interactions. Since then, attention-based algorithms \cite{tao2022sminet, wang2023statrl, wu2019pd, chen2022multi}
%
%
have been explored to learn representative items from historical interactions. Furthermore, attempts \cite{chang2021sequential, xia2022self, zhang2022dynamic, xia2022hypergraph} to incorporate sequential and spatial representations gained inspiring results. Another attempt has been made to improve training strategies, such as predicting masked items \cite{sun2019bert4rec} and intention disentanglement \cite{ma2020disentangled}.  More recently, contrastive learning (CL) draws intensive attention to derive self-supervision signals from user behavior sequences \cite{chen2022intent, xie2022contrastive}. 
However, they neglect that a user may still provide negative feedback on an item that significantly diverges from the user's preferences.

\noindent
\textbf{Graph Neural Network for Recommendation.}
Many efforts \cite{wang2019kgat, he2020lightgcn, wu2021self, yu2022graph, peng2022less, wang2022towards, wang2021learning} in GCNs have been made to capture high-order signals from the user\textendash item interactive graph. Several authors perform theoretical strategies to enhance GCNs, including pre-training GCNs \cite{hao2021pre}, low-pass collaborative filter networks~\cite{yu2022low}, multi-view intent disentangle graph networks \cite{zhao2022multi}. 
Furthermore, contrastive augmentations uniformize graph embeddings by ensuring the consistency between representations derived from contrastive graph views~\cite{yu2021socially, yu2023xsimgcl, lin2022improving, cai2022lightgcl}.
These approaches are further improved to mine rich information against data sparsity by stacking Transformers behind GCN layers \cite{xia2022hypergraph, xia2022self, wang2023eedn}.
%
%
However, few of them address the different feedback for the same item in GCN that results in defective node representations.


\noindent
\textbf{Negative Feedback Learning for Recommendation.}
Recent work exploits interactive events that provide negative feedback, such as low ratings \cite{huang2023negative, wang2021interactive}, dislikes \cite{zhao2018recommendations, wu2020neural, gong2022positive}, or skipped content \cite{huang2023negative, gong2022positive}, to create negative representations for recommendations. They treat negative and positive representations independently, failing to capture behavioral features across users' interacted items.
Some attempts exclude negative instances (e.g., ratings below 4 \cite{yu2021socially, yu2020enhancing, yu2022graph}) or regard all interaction records as positive feedback. They either make the task deviate from real-world scenarios or lose feedback features, resulting in a performance drop. 

\section{Preliminaries}

\subsection{Definitions}

\noindent
\textbf{(User-Item Feedback Graph)}.
The user-item feedback graph $\mathcal{G}$ = ($\mathcal{U}$, $\mathcal{I}$, $\mathcal{E}$) shows the network of feedback correlations between users and items, where $\mathcal{U}$ and $\mathcal{I}$ denote a set of users and items, respectively. $\mathcal{E}$ = \{ $\zeta_{u,i}$ | $u \in \mathcal{U}$, $i \in \mathcal{I}$\} denotes the set of edges.
Each element $\zeta_{u,i}$ equals 1 if user $u$ provides positive feedback for item $i$, -1 if user $u$ gives negative feedback for item $i$, and 0 otherwise.

\noindent
\textbf{(Item Recommendation)}.
Given a user historical interaction set $\mathcal{S}$, where $\mathcal{S}^{(u)} = \{i_{k}\}_{k=1}^{|\mathcal{S}^{(u)}|}$ denotes the item sequence of user $u$, and $|\mathcal{S}^{(u)}|$ refers to the number of items in $\mathcal{S}^{(u)}$, the goal is to generate a list of candidate items with the probabilities.                                                             

\subsection{Hawkes Process}

The basic concept of the Hawkes Process \cite{hawkes1971spectra} is to model the impact of past events on the occurrence of the current event, and the influence decreases over time. The original proposal suggests using exponential and power-law functions for the intensity function. Thereafter, due to the capacity of recurrent neural networks (RNNs) in modeling hidden states of sequential events, \citet{mei2017neural} propose a neural Hawkes process (NHP) to replace the intensity function with the learning process of RNNs. More recently, since the Transformer \cite{vaswani2017attention} can capture long-range dependencies between every two events in the sequence, \citet{zuo2020transformer} propose THP that adopts the self-attention mechanism as the core of its intensity function. Motivated by this, we utilize the THP to infer the feedback sentiment polarity of the next interactions, to capture underlying sentiment towards users' interacted items.


\subsection{Enhanced HGC with Feedback}
HGCs have been extensively employed to mine user\textendash item collaborative signals in recommendations that go beyond pairwise user-item connections \cite{ji2020dual, yu2021self, xia2022hypergraph, xia2022self, yan2023spatio} since each hyperedge in the hypergraph can connect more than two nodes~\cite{yu2019spectrum, yu2020graph}.
They typically regard nodes equally weighted in each hyperedge to perform convolution, while neglecting that various users may give opposite feedback for the same items. 
 As shown in Fig. \ref{fig:motivation} (b), the feedback sentiment of items $i_3$ and $i_4$ given by user $u_1$ are both positive, indicating that the HGC operator should pass information from $i_3$ to $i_4$. In contrast, there should be fewer messages from $i_6$ to $i_4$ as user $u_2$ gives different negative feedback on these two items.
%
%
 To address the issue, we introduce a two-phase HGC approach that leverages feedback relations between users and items to perform more efficient information propagation during convolutions. 


\section{Methodology}
\begin{figure*}[h]
\begin{minipage}{\linewidth}
\center{
\includegraphics[width=.83\linewidth]{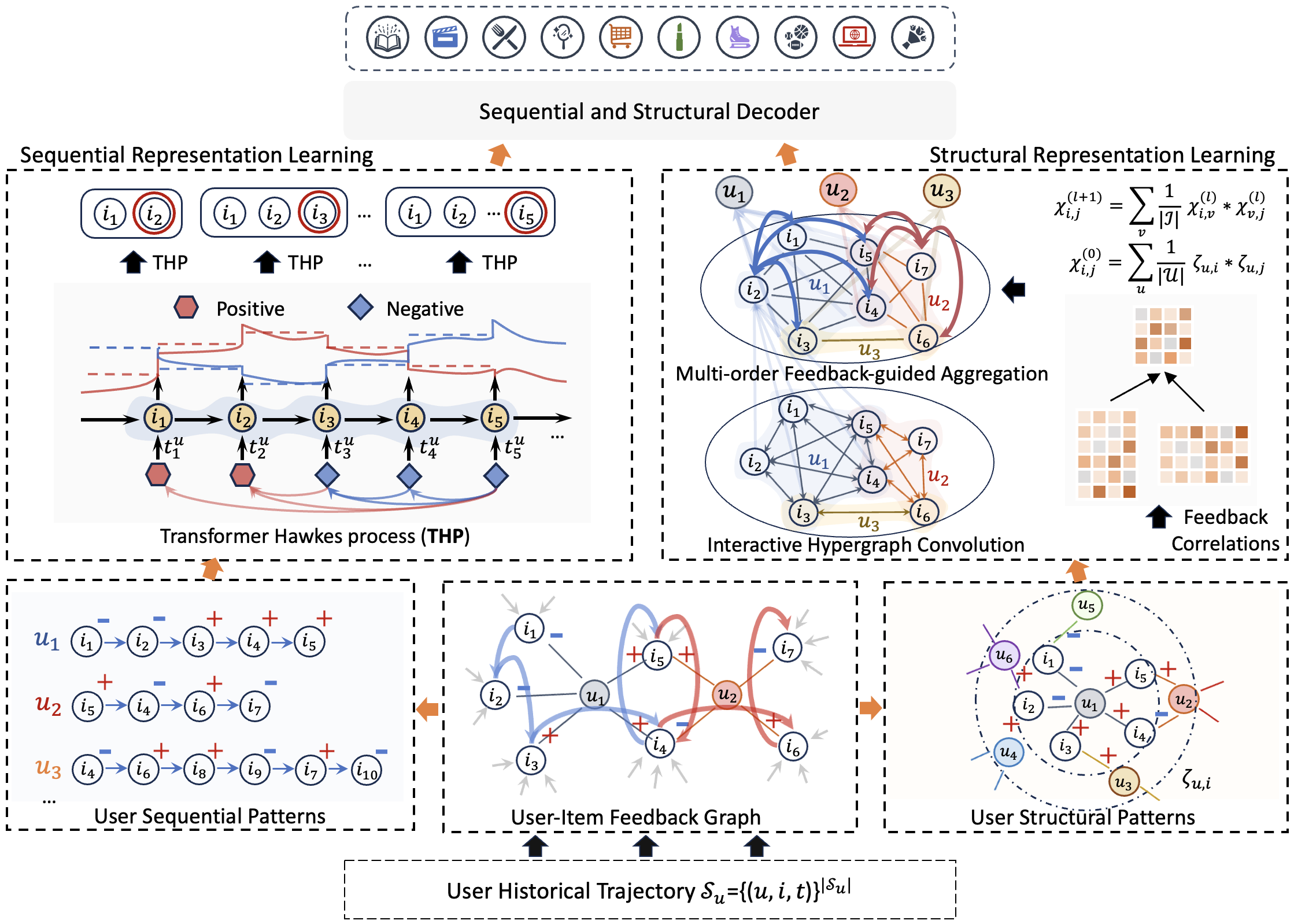}
  \caption{Architecture of our NFARec framework. NFARec learns negative feedback in both sequential and structural patterns.}
  \label{fig:framework}
}
\end{minipage}
\end{figure*}

The framework of NFARec is illustrated in Fig.~\ref{fig:framework}, which consists of sequential and structural representation learning. 
%


\subsection{Sequential Representation Learning}


Given a matrix $\mathcal{V} \in \mathbb{R}^{|\mathcal{I}| \times d_m}$ representing item embeddings, where $d_m$ is the dimension size, the embedding $v_i$ for an item $i$ is obtained by using the one-hot vector $\mathbf{v}^{'}_i \in \mathbb{R}^{\lvert \mathcal{I} \rvert}$ as a slicing operation that selects corresponding item embedding vectors on $\mathcal{V}$: $v_i = \mathcal{V}\mathbf{v}^{'}_i$.
%
%
%
%

Since each of the past interactions has an impact on the current interaction occurrence, we employ the Transformer blocks to capture long-term sequential dependencies through the self-attention scores as shown in the THP of Fig. \ref{fig:framework}. 
Considering $\mathcal{T}_t=\{(t_j, z_j): t_j<t\}$ as a user's historical interactions up to time $t$ with the feedback sentiment polarity $z_j \in \{positive, negative\}$, we feed the interaction sequence embedding $\mathcal{V}[\mathcal{T}_t] = \mathcal{V} \cdot Concat([\mathbf{v}^{'}_1, \mathbf{v}^{'}_2, ..., \mathbf{v}^{'}_t])$ as the input, $\mathbf{e}_t^{(0)}$ into the first layer of the Transformer. Thus, the representation $\hat{\mathbf{e}}_t^{(l)}$ of historical interactions at time $t$ in the $l$-th layer is obtained by the following equation:
%
\begin{equation}
\begin{aligned}
\label{eq:sa}
\mathbf{e}_t^{(l+1)} =  \alpha_1 \left (\sum_{j \in \mathcal{T}_t} \frac{ \mathrm{Mask}(\mathbf{e}_t^{(l)} \mathbf{w}^{q}_{*,t}  (\mathbf{e}_j^{(l)} \mathbf{w}^{k}_{*,j})^\top)}{\sqrt{d_k}}  \right ) \mathbf{e}_t^{(l)} \mathbf{w}^{v}_{*,i}, 
\end{aligned}
\end{equation}
where $\alpha_1$ denotes softmax activation function and $\mathbf{W}^{q}$, $\mathbf{W}^{k}$ and $\mathbf{W}^{v}$ are three weight matrices. 
The function $\mathrm{Mask}(\cdot)$ refers to a masking operation \cite{vaswani2017attention, zuo2020transformer} using an upper triangular matrix. This ensures that $\mathbf{e}_t$ is informed by $\mathbf{e}_{t-1}$ while avoiding $\mathbf{e}_{t-1}$ from peeking into future interactions.

After stacking the number of $L_1$ Transformer layers, an average-pooling operation, and an L2 norm normalization, we obtain the hidden representation $\mathbf{h}(t_j)=\mathbf{H}_{j,*}$ for each interaction $(t_j,z_j)$ in the sequence. The user sequential representations are obtained by considering the hidden states of users' historical interactions:
\begin{equation}
e^{\mathcal{S}}_u = \mathrm{SumPooling}(\{\mathbf{h}(t_j) : j \in |\mathcal{T}_t|\}).
\end{equation}




To further enhance user sequential representations by ensuring sentiment consistency among interacted items, we introduce an auxiliary task, i.e., predicting the feedback sentiment polarity of the next interaction. Specifically, let $\lambda(t|\mathcal{T}_t)$ be the conditional intensity function to model the influence of the past interactions up to time $t$. It accumulates the influences of both types of feedback in the past interactions as follows:
\begin{equation}
\lambda(t|\mathcal{T}_t) = \sum_z^Z \lambda_z (t|\mathcal{T}_t),
\end{equation}

\noindent
where the intensity ($\lambda_z(t|\mathcal{T}_t)$) at time $t$ is determined by the historical hidden state $\mathbf{h}(t_j)$, along with the current state denoted as $\alpha_z \frac{t-t_j}{t_j}$ (where $t_j$ is the timestamp of the last interaction before $t$):
\begin{equation}
\lambda_z(t|\mathcal{T}_t) = \mathrm{softplus} \left (\alpha_z \frac{t-t_j}{t_j}+\mathrm{MLP}(\mathbf{h}(t_j)) \right ),
\end{equation}

\noindent
where $\mathrm{softplus}(x)=\beta_z \log(1+\mathrm{exp(x/\beta_z))}$ is the $\mathrm{softplus}$ function with a trainable softness parameter $\beta_z$ and $\alpha_z$ indicates a learnable parameter.
The goal of the task is to maximize the intensity of the feedback sentiment polarity for every next interaction, i.e., $j \in |\mathcal{T}_t|$, which is given by:
\begin{equation}
\hat{z}_{j+1} = \underset{z}{\mathrm{argmax}} \ \frac{\lambda_z(t_{j+1}|\mathcal{T}_t)}{\lambda(t_{j+1}|\mathcal{T}_t)}.
\end{equation}

It is noteworthy that the hidden representation $\mathbf{h}(t)$ by NFARec is enhanced throughout the auxiliary task by predicting the feedback polarity of every next item. 
As such, NFARec explores the sentiment relatedness among the items in the interaction sequences. 
This enables NFARec to learn to predict items that the user will interact with, even if the user may not necessarily prefer them. 
%

\subsection{Structural Representation Learning}
Motivated by HGCs \cite{xia2022hypergraph, xia2022self, yin2022dynamic, yan2023spatio, wang2023eedn}, we propose a two-parse HGC paradigm to exploit the rich feedback features. As illustrated in Fig. \ref{fig:framework}, the paradigm includes an interactive hypergraph convolution and a multi-order feedback-aware aggregation. 

\noindent
\textbf{Interactive Hypergraph Convolution.}
Following the works~\cite{he2020lightgcn, wang2023eedn} that simplify GCNs and make them more appropriate for recommendations, we removed the global weight matrix and added a linear feature transformation ($\mathbf{W}_1 \in \mathbb{R}^{d_m \times d_m}$). 
Given the global item embeddings $\mathcal{V}$ as the input $\Lambda^{(0)}$ of the first layer and the user hyperedge $\mathcal{H}^{(u)} \in \mathbb{R}^{1 \times |\mathcal{I}|}$, the matrix equivalence form of informative diffusion in the HGC is as follows:
\begin{equation}
\Lambda^{(l+1)}= \mathcal{H}^{(u)}\hat{\mathcal{A}} {\mathcal{H}^{(u)}}^\top \alpha_2(\Lambda^{(l)} \mathbf{W}_1), \\
\end{equation}

\noindent
where $\Lambda^{(l)} \in \mathbb{R}^{|\mathcal{I}| \times d_m}$ indicates the output of the $l$-th HGC layer, $\alpha_2$ is the ELU activation function, and $\hat{\mathcal{A}}$ denotes the symmetrically normalized adjacency matrix:
\begin{equation}
\hat{\mathcal{A}} = \mathbf{D}^{-\frac{1}{2}} \mathcal{A} \mathbf{D}^{-\frac{1}{2}},  \\
\end{equation}
where $\mathcal{A} \in \mathbb{R}^{|\mathcal{I}| \times |\mathcal{I}|}$ denotes the global adjacent matrix for the item-item graph, and $a_{ij} = 1$, if item $i$ and $j$ have interacted with the same user, otherwise 0. $\mathbf{D} \in \mathbb{R}^{|\mathcal{I}| \times |\mathcal{I}|}$ is a diagonal matrix in which each entry $d_{ij}$ denotes the number of nonzero entries in the $i$-th row vector of the adjacency matrix $\mathcal{A}$.
 
After repeating the number of $L_2$ HGCs and an average pooling operation, the user structural representations are obtained by aggregating the representations of users' interacted items:
\begin{equation}
e^{\mathcal{H}_1}_u =  \mathrm{AvgPooling}(\Lambda^{(L_2)}).
\end{equation}

\begin{figure}
  \includegraphics[width=.95\linewidth]{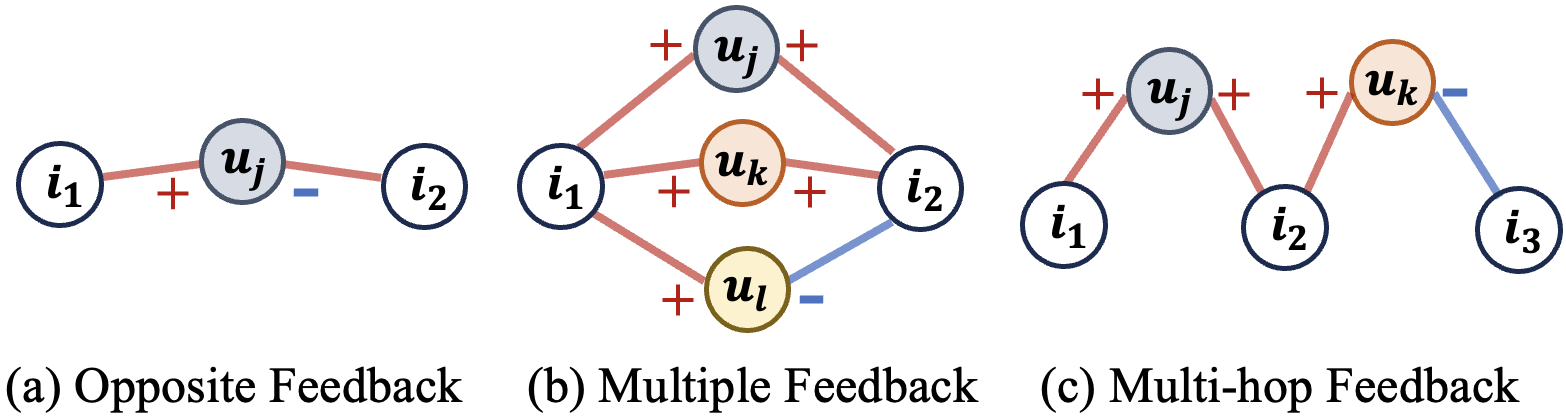}
  \caption{Illustration of various feedback.  }
  \label{fig:various_feedback}
\end{figure}

\noindent
\textbf{Feedback-aware Aggregation.} To spread the information through an optimal path during convolution, we exploit user feedback to enhance HGC operators. There are two common scenarios among adjacent nodes, which are illustrated in Fig. \ref{fig:various_feedback}: (a) opposite feedback, where user $u_j$ provides positive and negative feedback on $i_1$ and $i_2$, respectively; (b) multiple feedback, where three users — $u_j$, $u_k$, and $u_l$ — offer varying feedback on $i_1$ and $i_2$. Hence, we construct a feedback correlation matrix $\mathcal{Z} \in \mathbb{R}^{|\mathcal{U}| \times |\mathcal{I}|}$. Each element $\zeta_{i,j}$ of $\mathcal{Z}$ is set to 1 if user $u_i$ provides positive feedback for item $v_j$, -1 if user $u_i$ gives negative feedback for item $v_j$, and 0 otherwise. Subsequently, we create the 1st-order feedback correlation matrix $\mathcal{X}^{(0)}  \in \mathbb{R}^{|\mathcal{I}| \times |\mathcal{I}|}$ by incorporating feedback among users interacting with the same items:
\begin{equation}
\mathcal{X}^{(0)} = 
\begin{bmatrix}
    \chi^{(0)}_{1, 1} & \chi^{(0)}_{1, 2} & \cdots & \chi^{(0)}_{1, |\mathcal{I}|}\\
    \chi^{(0)}_{2, 1} & \chi^{(0)}_{2, 2} & \cdots & \chi^{(0)}_{2, |\mathcal{I}|}\\
    \vdots & \vdots & \ddots & \vdots \\
    \chi^{(0)}_{|\mathcal{I}|, 1} & \chi^{(0)}_{|\mathcal{I}|, 2} & \cdots & \chi^{(0)}_{|\mathcal{I}|, |\mathcal{I}|}\\
\end{bmatrix}, \chi^{(0)}_{i, j} =\sum_{u \in \mathcal{U}}  \frac{\zeta_{i, u}\cdot \zeta_{u, j}}{|\mathcal{U}|}. \\
\end{equation}

To model the multi-hop feedback correlations, such as the relationship between item $i_1$ and $i_3$ in Fig. \ref{fig:various_feedback} (c), we distill the high-order relationships from the 1st-order feedback correlations. The correlation matrices denoted by $\mathcal{X}^{(l)}$ are computed layer by layer:
%
%
\begin{equation}
\mathcal{X}^{(l+1)} = 
\begin{bmatrix}
    \chi^{(l+1)}_{1, 1} & \chi^{(l+1)}_{1, 2} & \cdots & \chi^{(l+1)}_{1, |\mathcal{I}|}\\
    \chi^{(l+1)}_{2, 1} & \chi^{(l+1)}_{2, 2} & \cdots & \chi^{(l+1)}_{2, |\mathcal{I}|}\\
    \vdots & \vdots & \ddots & \vdots \\
    \chi^{(l+1)}_{|\mathcal{I}|, 1} & \chi^{(l+1)}_{|\mathcal{I}|, 2} & \cdots & \chi^{(l+1)}_{|\mathcal{I}|, |\mathcal{I}|}\\
\end{bmatrix}, \chi^{(l+1)}_{i, j} =\sum_{v \in \mathcal{I}}  \frac{\chi^{(l)}_{i, v}\cdot \chi^{(l)}_{v, j}}{|\mathcal{I}|}. \\
\end{equation}

To integrate the multi-order correlations across different orders, we define the correlation matrix by $\hat{\mathcal{X}} = (\mathcal{X}^{(0)} + \mathcal{X}^{(1)} + \cdots)$. Similar to Eq. (\ref{eq:sa}), we utilize the masking technique to avoid message-passing between conflicting nodes. Specifically, if an element in the mask is below 0, it is masked; otherwise, it remains unchanged. The feedback-aware user representation is given as follows:
%
\begin{equation}
e^{\mathcal{H}_2}_u = \mathrm{AvgPooling}(\mathcal{H}^{(u)}\mathrm{Mask}(\hat{\mathcal{X}}) \Lambda^{(L_2)}). \\
\end{equation}


Recall that the HGC paradigm models the interactions between users and items in the first stage, and utilizes the feedback to guide HGC operators in the second stage, enabling the NFARec to learn both behavioral features and user feedback for recommendations.

\subsection{Sequential and Structural Decoder}

We employ the inner products between user sequential and structural embeddings, i.e., $e^{\mathcal{S}}_u$ and $e^{\mathcal{H}_1}_u$, and the item embeddings, i.e., $\mathcal{V}$, to infer their preferences in candidate items. In addition, the model leverages the feedback-aware structural feature, $e^{\mathcal{H}_2}_u$, to enhance its robustness to user feedback. We define the sequential, structural, and feedback-aware structural features as Seq, Gra$_1$, and Gra$_2$, respectively. The final rating scores over all candidate items are given by:
%
\begin{equation}
\mathbf{r}_{u,*} =  \alpha_3(\underbrace{e^{\mathcal{S}}_u {\mathcal{V}}^\top}_{\text{Seq}} + \underbrace{ e^{\mathcal{H}_1}_u {\mathcal{V}}^\top}_{\text{Gra}_1} +  \underbrace{\delta \mathrm{MLP}(e^{\mathcal{H}_2}_u)}_{\text{Gra}_2}),\\
\end{equation}

\noindent
where $\mathrm{MLP}$ denotes a linear transformer to decode Gra$_2$ and $\alpha_3$ indicates the $\mathrm{tanh}$ activation function. $\delta$ refers to the hyperparameter to assign weight to the third term.


\subsection{Model Optimization}
The NFARec consists of a main task, a recommendation, and an auxiliary task, the next feedback sentiment polarity prediction.

\noindent
\textbf{Recommendation.} We adopt a weighted multi-label cross-entropy loss~\cite{wang2023eedn} as the main objective function:
\begin{equation}
\label{eq:cross_entropy}
  \mathcal{L}_{main} = \sum_{(u \in \mathcal{U})}  \sum_{(i \in \mathcal{I})} - c_{u,i} * \gamma_{i}^{(u)} \log(\alpha_4(\hat{\mathbf{r}}_{u,i})),
\end{equation}

\noindent
where $\mathbf{\gamma}^{(u)} \in \mathbb{R}^{|\mathcal{I}|}$ represents a label vector, in which the element equals 1 if the corresponding item is a ground-truth candidate; otherwise, 0. $\alpha_4$ denotes the sigmoid activation function. $c_{u,i}$ is set to $\beta_1$ when the user has interacted with the item, $\beta_2$ when the user will interact with the item, and 0 otherwise. 

\noindent
\textbf{Next Feedback Sentiment Polarity Prediction.} Given user $u$'s interaction sequence $\mathcal{S}_u$ with a time interval of $[t_1, t_{|\mathcal{S}_u|}]$, the log-likelihood for the task is given as follows:
\begin{equation}
\begin{aligned}
  \mathcal{L}_{auxi} &= \sum_{(u \in \mathcal{U})} ( \underbrace{\sum_{j=1}^{|\mathcal{S}_u|} \log \lambda(t_j|\mathcal{T}_t)}_{\text{interactive log-likelihood}}
  -  \underbrace{\vphantom{\sum_{j=1}^{|\mathcal{S}_u|}}\int^{t_{|\mathcal{S}_u|}}_{t_1} \lambda(t|\mathcal{T}_t) dt }_{\text{continuous log-likelihood}} ).
\end{aligned}
\end{equation}

The task is learned by maximizing interactive and continuous log-likelihood. The final multi-task objective loss is given by:
%
\begin{equation}
\begin{aligned}
  \mathcal{L}_{final} = \mathcal{L}_{main} + \delta_2 \mathcal{L}_{auxi}, \\
\end{aligned}
\end{equation}

\noindent
where $\delta_2$ is the hyperparameter for the auxiliary task.

To make the continuous log-likelihood, i.e., $\Delta=\int_{t_i}^{t_{|\mathcal{S}_u|}} \lambda(t|\mathcal{T}_t) dt$, trainable by back-propagation, we exploit Monte Carlo integration (MCI) \cite{robert1999monte} following Zuo et al.'s work \cite{zuo2020transformer}, which leverages random sampling to estimate definite integrals, to approximate $\Delta$:
\begin{equation}
\begin{aligned}
\hat{\Delta}_{MCI} = \sum_{j=2}^{|S_u|} (t_j-t_{j-1}) \left (\frac{1}{N}\sum_{k=1}^{N} \lambda(x_k) \right),
\end{aligned}
\end{equation}

\noindent
where $N$ represents the number of sampling points in MCI, and $x_k \sim \mathrm{uniform}(t_{j-1}, t_j)$ is a uniform distribution such that $\lambda(x_k)$ can be calculated by back-propagation through the NFARec and the unbiased estimation $\mathbb{E}[\hat{\Delta}_{MCI}]$ is equal to $\Delta$.

\section{Experiment}

\label{sec:experiment_setup}
\begin{table}
  \caption{Statistics of datasets. "\#Avg." denotes the average count of users' interactions. ``Perc.(\#Pos/\#Neg)'' refers to the percentage of positive and negative samples.}
  \label{tab:statistics}
\resizebox{\linewidth}{!}{ 
  \begin{tabular}{cccccc}
    \toprule
    {Dataset} &{\#Users} &{\#Items} &{\#Interactions}& Perc.(\#Pos/\#Neg) & \#Avg.\\
    \midrule
    {Yelp 2023} &48,993& 34,298& 3,415,569& 72.4\%/27.6\%& 39.5 \\
    {Beauty} & 22,363& 12,101& 198,502& 80.2\%/19.8\%& 8.9\\
    {Book} &19,804& 22,086& 1,273,679& 80.6\%/19.4\%& 64.3 \\
    {Recipes} &7,452 & 12,911& 311,389& 82.5\%/17.5\%& 41.8 \\
    {MovieLens} &6,041& 3,955& 1,000,209& 73.5\%/26.5\%& 165.6 \\
  \bottomrule
\end{tabular}
}
\end{table}

\setlength{\tabcolsep}{1.1mm}{
\begin{table*}
  \caption{Performance comparison of recommender models. Bold: Best, \uline{underline}: Second best. ``*'' indicates that the improvement is statistically significant (p-value < 0.01) in the 10-trial T-test.}
  \label{tab:comparison}
\resizebox{1\linewidth}{!}{ 
  \begin{tabular}{c|c|cc|ccc|ccc|c|ccc|c|rc}
    \toprule
    \multirow {2}{*}{Dataset } & \multirow {2}{*}{Metric } &\multicolumn{2}{c|}{CF-based}  
    &\multicolumn{3}{c|}{GCN-based} & \multicolumn {3}{c|}{GCL-based } & Dif-based & \multicolumn {3}{c|}{S\&G-based }& \multirow {2}{*}{\textbf{Ours}} & \multirow {2}{*}{Improv.} & \multirow {2}{*}{p-value} 
    \\ 
     &  & \vspace{0.1em} \small{DirectAU}& \ \ InvCF\ \ & \small{LightGCN}& \ \ SGL\ \ & \ \  HCCF\ \ &\ \  NCL \ \ & \small{LightGCL}  & \small{XSimGCL}& DiffRec & \ \ SHT \ \  & AutoCF& \ \ EEDN\ \ & & \\
    \midrule
    \midrule
    \multirow {6}{*}{Yelp 2023} 
    & Recall@5  & 0.0316& 0.0321& 0.0344& 0.0366& 0.0299& 0.0360 & 0.0341& 0.0376& 0.0323& 0.0340& 0.0391& \uline{0.0426}& \textbf{0.0449} & +5.6\% *  & 1.4e-07\\
    & NDCG@5    & 0.0497& 0.0607& 0.0540& 0.0551& 0.0426& 0.0535 & 0.0479& 0.0566& 0.0487& 0.0619& 0.0614& \uline{0.0668}& \textbf{0.0741} & +10.9\% * & 2.4e-09\\
    & Recall@10 & 0.0517& 0.0464& 0.0564& 0.0581& 0.0403& 0.0572 & 0.0502& 0.0590& 0.0529& 0.0476& 0.0535& \uline{0.0672}& \textbf{0.0707} & +5.2\% *  & 2.2e-06\\
    & NDCG@10   & 0.0507& 0.0644& 0.0587& 0.0603& 0.0457& 0.0571 & 0.0498& 0.0595& 0.0548& 0.0673& 0.0704& \uline{0.0733}& \textbf{0.0778} & +6.1\% *  & 4.9e-07\\
    & Recall@20 & 0.0722& 0.0798& 0.0752& 0.0721& 0.0518& 0.0795 & 0.0675& 0.0857& 0.0809& 0.0796& 0.0939& \uline{0.1031}& \textbf{0.1074} & +4.2\% *  & 6.1e-06\\
    & NDCG@20   & 0.0714& 0.0795& 0.0715& 0.0740& 0.0526& 0.0743 & 0.0661& 0.0756& 0.0713& 0.0785& 0.0810& \uline{0.0858}& \textbf{0.0894} & +4.1\% *  & 2.2e-06\\
    \midrule
    \multirow {6}{*}{MovieLens} 
    & Recall@5  & 0.0726& 0.0728& 0.0603& 0.0665& 0.0603& 0.0673 & 0.0690& 0.0726& 0.0642& 0.0698& 0.0691& \uline{0.0737}& \textbf{0.0789} & +7.1\% *  & 1.8e-08\\
    & NDCG@5    & 0.2969& 0.3144& 0.2438& 0.2767& 0.2490& 0.2579 & 0.2556& 0.3165& 0.2710& 0.2822& 0.3084& \uline{0.3222}& \textbf{0.3716} & +15.3\% * & 1.9e-12\\
    & Recall@10 & 0.1044& 0.1085& 0.0863& 0.1069& 0.0967& 0.1029 & 0.1046& 0.1037& 0.0923& 0.1072& 0.0991& \uline{0.1136}& \textbf{0.1213} & +6.7\% *  & 8.3e-06\\
    & NDCG@10   & 0.2819& 0.2831& 0.2476& 0.2829& 0.2539& 0.2497 & 0.2494& 0.2754& 0.2670& 0.2739& 0.2729& \uline{0.2878}& \textbf{0.3314} & +15.1\% * & 3.03-13\\
    & Recall@20 & 0.1504& 0.1481& 0.1107& 0.1432& 0.1287& 0.1298 & 0.1313& 0.1426& 0.1204& 0.1509& 0.1470& \uline{0.1628}& \textbf{0.1736} & +6.6\% *  & 3.5e-07\\
    & NDCG@20   & 0.2592& 0.2537& 0.2047& 0.2541& 0.2312& 0.2479 & 0.2540& 0.2481& 0.2284& 0.2400& 0.2452& \uline{0.2631}& \textbf{0.3007} & +14.2\% * & 5.6e-10\\
    \midrule
    \multirow {6}{*}{Recipes} 
    & Recall@5  & 0.0235& 0.0167& 0.0208& 0.0265& 0.0224& 0.0207 & 0.0164& 0.0253& 0.0193& 0.0249& 0.0162& \uline{0.0269} & \textbf{0.0292} & +8.6\% * & 1.6e-09\\
    & NDCG@5    & 0.0359& 0.0352& 0.0283& 0.0365& 0.0330& 0.0267 & 0.0257& 0.0389& 0.0283& 0.0377& 0.0264& \uline{0.0395} & \textbf{0.0418} & +9.4\% * & 1.5e-08\\
    & Recall@10 & 0.0426& 0.0266& 0.0333& 0.0406& 0.0340& 0.0330 & 0.0296& \uline{0.0419}& 0.0324& 0.0405& 0.0273& 0.0401 & \textbf{0.0442} & +5.5\% * & 3.6e-06\\
    & NDCG@10   & 0.0429& 0.0254& 0.0304& 0.0382& 0.0332& 0.0293 & 0.0275& \uline{0.0435}& 0.0305& 0.0412& 0.0237& 0.0429 & \textbf{0.0459} & +5.5\% * & 1.4e-06\\
    & Recall@20 & 0.0595& 0.0376& 0.0399& 0.0566& 0.0386& 0.0402 & 0.0360& \uline{0.0595}& 0.0416& 0.0560& 0.0331& 0.0592 & \textbf{0.0609} & +2.4\% * & 1.4e-03\\
    & NDCG@20   & 0.0476& 0.0227& 0.0329& 0.0465& 0.0346& 0.0319 & 0.0299& \uline{0.0481}& 0.0392& 0.0406& 0.0186& 0.0475 & \textbf{0.0492} & +2.3\% * & 2.3e-04\\
    \midrule
    \multirow {6}{*}{Books} 
    & Recall@5  & 0.0392& 0.0301& 0.0329& 0.0338& 0.0252& 0.0359 & 0.0275& 0.0440& 0.0254& 0.0317& 0.0329& \uline{0.0473}& \textbf{0.0489} & +3.4\% * & 1.7e-04\\
    & NDCG@5    & 0.0973& 0.0827& 0.0905& 0.0874& 0.0524& 0.0912 & 0.0723& 0.1029& 0.0654& 0.0834& 0.0890& \uline{0.1082}& \textbf{0.1139} & +5.3\% * & 1.0e-06\\
    & Recall@10 & 0.0696& 0.0590& 0.0655& 0.0693& 0.0474& 0.0667 & 0.0513& 0.0712& 0.0484& 0.0680& 0.0516& \uline{0.0729}& \textbf{0.0753} & +3.3\% * & 1.1e-05\\
    & NDCG@10   & 0.0988& 0.0804& 0.0956& 0.0896& 0.0635& 0.0969 & 0.0715& 0.0990& 0.0632& 0.0913& 0.0727& \uline{0.1011}& \textbf{0.1065} & +5.3\% * & 3.1e-07\\
    & Recall@20 & 0.1029& 0.0854& 0.0994& 0.0981& 0.0670& 0.0971 & 0.0672& 0.1037& 0.0647& 0.0828& 0.0762& \uline{0.1075}& \textbf{0.1112} & +3.4\% * & 7.8e-05\\
    & NDCG@20   & 0.1024& 0.0792& 0.1007& 0.0992& 0.0812& 0.0943 & 0.0645& 0.1033& 0.0560& 0.0799& 0.0730& \uline{0.1099}& \textbf{0.1141} & +3.9\% * & 1.1e-08\\
    \midrule
    \multirow {6}{*}{Beauty} 
    & Recall@5  & 0.0518& 0.0530& 0.0489& 0.0522& 0.0367& 0.0521 & 0.0468& 0.0548& 0.0479& 0.0520& 0.0490& \uline{0.0548}& \textbf{0.0585} & +6.8\% * & 5.7e-07\\
    & NDCG@5    & 0.0447& 0.0476& 0.0414& 0.0467& 0.0344& 0.0453 & 0.0417& 0.0486& 0.0402& 0.0415& 0.0437& \uline{0.0476}& \textbf{0.0514} & +8.0\% * & 7.8e-07\\
    & Recall@10 & 0.0725& 0.0710& 0.0704& 0.0737& 0.0517& 0.0688 & 0.0615& 0.0733& 0.0657& 0.0719& 0.0759& \uline{0.0779}& \textbf{0.0811} & +4.1\% * & 1.2e-06\\
    & NDCG@10   & 0.0527& 0.0522& 0.0499& 0.0530& 0.0387& 0.0502 & 0.0488& 0.0557& 0.0476& 0.0513& 0.0542& \uline{0.0558}& \textbf{0.0596} & +6.8\% * & 8.1e-09\\
    & Recall@20 & 0.0908& 0.1022& 0.0879& 0.0939& 0.0672& 0.0839 & 0.0803& 0.1007& 0.0876& 0.0901& 0.1034& \uline{0.1054}& \textbf{0.1103} & +4.6\% * & 3.0e-05\\
    & NDCG@20   & 0.0614& 0.0613& 0.0570& 0.0607& 0.0413& 0.0529 & 0.0499& 0.0640& 0.0517& 0.0587& 0.0625& \uline{0.0648}& \textbf{0.0683} & +5.4\% * & 4.1e-05\\
    \bottomrule
  \end{tabular}
  }
\end{table*}
}

\subsection{Experimental Setup}
\noindent 
\textbf{Datasets and Evaluation Metrics.} To examine user feedback, we collected five public datasets with rating scores, i.e., Yelp 2023\footnote{https://www.yelp.com/dataset}, Beauty\footnote{https://www.kaggle.com/datasets/skillsmuggler/amazon-ratings/}, Books\footnote{https://jmcauley.ucsd.edu/data/amazon/links.html},
Recipes\footnote{https://www.kaggle.com/datasets/shuyangli94/food-com-recipes-and-user-interactions/}, and MovieLens \cite{lin2022improving}. Similar to previous works \cite{huang2023negative, wang2021interactive, yu2021socially, yu2020enhancing, yu2022graph}, we regard interactions with ratings below 4 as negative feedback; others are positive. 
We exclude users and items with fewer than $n$ interactions, where $n$ is 15, 10, 25, and 5 for Yelp 2023, Recipes, Books, and Beauty, respectively. The statistical details are summarized in
Table~\ref{tab:statistics}. Following
~\cite{wang2023eedn, yu2022graph}, we split each dataset into
training, validation, and test data with a ratio of 7:1:2. We employed the widely used Recall@$K$ (R@$K$) and NDCG@$K$ (N@$K$) with $K$ $\in$ \{5, 10, 20\} as metrics to evaluate performance.

\noindent
\textbf{Baselines.} We compared the NFARec with the following 12 SOTA recommender models which are classified into five groups:

\vspace{0.2em}
\noindent
- {\it \textbf{Collaborative Filtering (CF)-based method}}:

\textbf{DirectAU}~\cite{wang2022towards} optimizes alignment and uniformity of graph embeddings in the hypersphere. \textbf{InvCF}~\cite{zhang2023invariant} discovers disentangled the preference and popularity semantics for recommendations. 

\vspace{0.2em}
\noindent
- {\it \textbf{Graph Convolution Network (GCN)-based method}}:

\textbf{LightGCN}~\cite{he2020lightgcn} simplifies GCNs, making them more appropriate for recommendations. \textbf{SGL}~\cite{wu2021self} exploits a self-supervised task to reinforce node representation. \textbf{HCCF}~\cite{wu2021self} proposes hypergraph-enhanced cross-view contrastive learning for recommendations.


\vspace{0.2em}
\noindent
- {\it \textbf{Graph Contrastive Learning (GCL)-based method}}:

\textbf{NCL}~\cite{lin2022improving} regards users (or items) and their structural neighbors as contrastive pairs. \textbf{LightGCL}~\cite{cai2022lightgcl} utilizes singular value decomposition for refining contrastive learning. \textbf{XSimGCL}~\cite{yu2022graph} adds uniform noise to the embedding space to create contrastive views.


\vspace{0.2em}
\noindent
- {\it \textbf{Diffusion (Dif)-based method}}:

\textbf{DiffRec}~\cite{wang2023diffusion} learns the generative process in a denoising manner from personalized information in user interactions.

\vspace{0.2em}
\noindent
- {\it \textbf{Sequential and Graph (S\&G)-based method}}:

\textbf{SHT}~\cite{xia2022self} explores the self-attention- and hypergraph-based collaborative relationships. \textbf{AutoCF}~\cite{autocf2023} exerts self-supervised learning to automatically augment data for recommendations. \textbf{EEDN}~\cite{lin2022improving} exploits rich latent features between users, items, and interactions between users and items.


\noindent
\textbf{Implementation and Parameter Settings.}
The best parameters of the FNGRec were sampled as follows: $\beta_1$,
$\beta_2$, $\delta_1$, and $\delta_2$ were set to 0.12, 1.49, 1.2, and 1e-3 for Yelp 2023, 1.47, 3.99, 1.2, and 0.5 for MovieLens, 0.12, 3.81, 1.0 and 1e-5 for Recipes, 0.25, 3.53, 1.2, and 1e-5 for Books, and 0.62, 3.74, 1.2, and 1e-3 for Beauty, respectively. $d_m$ was 1,024. The numbers $L_1$ and $L_2$ of Transformer layers and HGC layers were 1. 
These hyperparameters were tuned using Optuna\footnote{https://github.com/pfnet/optuna}. 
The parameters for the baselines were tuned to attain the best performance or set as proposed by the authors.
We implemented our NFARec and experimented with Pytorch on Nvidia GeForce RTX 3090 (24GB memory).

\subsection{Overall Performance}
Table \ref{tab:comparison} shows the performance comparison between NFARec and SOTA baselines on five datasets. The results highlight the effectiveness of NFARec in making the most of negative feedback for recommendations. 
Specifically, the NFARec surpasses the runner-ups XSimGCL and EEDN by 4.2 $\sim$ 5.6\% in Recall@$k$ and 4.1 $\sim$ 10.9\% in NDCG@$k$ on the Yelp 2023 dataset, 6.6 $\sim$ 7.1\% in Recall@$k$ and 14.2 $\sim$ 15.3\% in NDCG@$k$ on the MovieLens dataset, 2.4 $\sim$ 8.6\% in Recall@$k$ and 2.3 $\sim$ 9.4\% in NDCG@$k$ on the Recipes dataset, 3.3 $\sim$ 3.4\% in Recall@$k$ and 3.9 $\sim$ 5.3\% in NDCG@$k$ on the Books dataset, and 4.1 $\sim$ 6.8\% in Recall@$k$ and 5.4 $\sim$ 8.0\% in NDCG@$k$ on the Beauty dataset. Table. \ref{tab:comparison} also promotes the following observations and insights:


(1) The S\&G-based methods are competitive among baselines, which indicates the effectiveness of both sequential and graph-based features to improve performance. Among them, EEDN achieves the highest performance across all datasets, suggesting that mining implicit features is beneficial to mitigate exposure bias for recommendations. However, their underperformance compared to NFARec shows that they do not take into account negative feedback in modeling user-item interactions.

(2) The GCN- and GCL-based methods capture high-order interactive information between users and items via GCNs. However, they typically treat neighboring nodes as equal in GCN and perform convolution along the connections between nodes, leading to a performance drop. The algorithms SGL, NCL, and XSimGCL outperform others because of their effective graph augmentations on edges or nodes. 
The lower effectiveness of LightGCL may be due to its susceptibility to conflicting feedback from different users, but this aspect was overlooked.

(3) For the CF-based methods, DirectAU optimizes the perspective of alignment and uniformity in the hypersphere and InvCF disentangles representations of preference and popularity semantics to refine the invariant information. They are dedicated to promoting better user and item representations but neglect the underlying feedback correlations in the user-item interactive network.

(4) DiffRec utilizes generative diffusion models for personalized interactive information learning. Its suboptimal performance suggests potential explorations for integrating negative feedback into conditional diffusion models.

\subsection{Ablation Study}
We conducted an ablation experiment to examine the effectiveness of each module in NFARec. Table. \ref{tab:ablation} shows the result in NDCG@20. Overall, the sequential features contribute more than the structural features in NFARec. We have the following findings:

(1) The sequential features enhance performance across five datasets, with an improvement ranging from 3.5\% to 11.1\%. This shows the efficacy of the auxiliary task, which predicts the feedback sentiment polarity of the next interaction.


(2) On the Yelp 2023, Recipes, and Beauty datasets, the graph-based feature (Gra$_1$) is effective, but its influence is not drastically on the Books dataset and harms the performance of the MovieLens dataset. This indicates that in several real-world scenarios, features derived from an interactive graph, without addressing conflicting feedback, have limited or even negative effects on performance.

(3) The feedback-aware graph-based feature benefits the recommendation quality for all datasets. It supports our assertion that the enhanced HGC operators effectively convey information along the feedback-aware paths. 





\begin{table}
  \caption{\small{Ablation study. ``w/o X'' denotes the removed parts. ``Seq'', ``Gra$_1$'', and ``Gra$_2$'' indicate the sequential, graph-based, and feedback-aware graph-based features, respectively. Note that on MovieLens, the results of ``w/o Seq'' and ``w/o Gra$_2$'' are obtained without ``Gra$_1$''.}}
  \label{tab:ablation}
\resizebox{.85\linewidth}{!}{ 
  \begin{tabular}{c|c|c|c|c|c}
    \toprule
      Model &Yelp 2023 & \small{MovieLens} & Recipes  & Books  & Beauty \\
      \midrule
    \multirow {2}{*}{w/o Seq} 
             & 0.0862 & 0.2741 & 0.0460 & 0.1026 & 0.0654 \\
             & (\textcolor{red}{+3.7\%}) & (\textcolor{red}{+9.7\%}) & (\textcolor{red}{+7.0\%}) & (\textcolor{red}{+11.2\%}) & (\textcolor{red}{+4.4\%}) \\
      \midrule
    \multirow {2}{*}{w/o Gra$_1$} 
             & 0.0866 & \textbf{0.3007} & 0.0462 & 0.1133 & 0.0648  \\
             & (\textcolor{red}{+3.2\%}) & (\textcolor{blue}{-5.2\%}) & (\textcolor{red}{+6.5\%}) & (\textcolor{red}{+0.7\%}) & (\textcolor{red}{+5.4\%}) \\
      \midrule
    \multirow {2}{*}{w/o Gra$_2$} 
             & 0.0873 & 0.2865 & {0.0466} & {0.1094} & {0.0669}\\
             & (\textcolor{red}{+2.4\%}) & (\textcolor{red}{+5.0\%}) & (\textcolor{red}{+5.6\%}) & (\textcolor{red}{+4.3\%}) & (\textcolor{red}{+2.1\%}) \\
      \midrule
    Full & \textbf{0.0894} & 0.2851 & \textbf{0.0492} & \textbf{0.1141} & \textbf{0.0683} \\
  \bottomrule
\end{tabular}
}
\end{table}

\begin{table}
  \caption{\small{Comparison of sequential encoders. ``THP'' and ``NHP'' indicate the Transformer Hawkes process and neural Hawkes process, respectively. ``Masking'' refers to the masking technology in Eq. (\ref{eq:sa}). }}
  \label{tab:encoders}
\resizebox{.93\linewidth}{!}{ 
  \begin{tabular}{c|cc|cc|cc}
    \toprule
      \multirow {2}{*}{Encoder } &\multicolumn{2}{c|}{Yelp 2023} &\multicolumn{2}{c|}{MovieLens} &\multicolumn{2}{c}{Recipes} \\
      \cline{2-7}
       & \ \ R@20\ \  &\ \  N@20\ \  &\ \  R@20\ \  &\ \  N@20\ \  &\ \  R@20\ \  &\ \  N@20\ \  \\ 
      \midrule
    Transformer & \uline{0.0960}& 0.0777& \uline{0.1512}& \uline{0.2543}& 0.0555& 0.0407\\
    NHP & 0.0921& \uline{0.0793}& 0.1478& 0.2402& \uline{0.0587}& \uline{0.0480} \\
    THP & \textbf{0.1074}& \textbf{0.0894}& \textbf{0.1736}& \textbf{0.3007}& \textbf{0.0609}& \textbf{0.0492} \\
      \midrule
    w/o Masking & 0.1053& 0.0876& 0.1608& 0.2765& 0.0574& 0.0463\\
  \bottomrule
\end{tabular}
}
\end{table}

\label{sec:experiment_setup}
\begin{table}
  \caption{Comparison of feedback across various orders.}
  \label{tab:feedback_orders}
\resizebox{.93\linewidth}{!}{ 
  \begin{tabular}{c|c|ccccc}
    \toprule
            & \ \ Metric \ \ & \ \small{Yelp 2023} & \ \small{MovieLens} \ & Recipes \ &\ \ Books\ \  & \ Beauty\\
    \midrule
     \multirow {2}{*}{\textbf{1-Order} } & R@20& 0.1071 & 0.1725 & \textbf{0.0609} & \textbf{0.1112} & 0.1098\\
                                         & N@20& 0.0892 & 0.2980 & \textbf{0.0492} & \textbf{0.1141} & 0.0678\\
    \midrule
     \multirow {2}{*}{\textbf{2-Order} } & R@20& \textbf{0.1074} & \textbf{0.1736} & 0.0605 & 0.1111& \textbf{0.1103}\\
                                         & N@20& \textbf{0.0894} & \textbf{0.3007} & 0.0486 & 0.1139& \textbf{0.0683}\\
    \midrule
     \multirow {2}{*}{\textbf{3-Order} } & R@20& 0.1071 & 0.1713& 0.0602& 0.1105 & 0.1101\\
                                         & N@20& 0.0893 & 0.2961& 0.0482& 0.1133 & 0.0681\\
    \midrule
     \multirow {2}{*}{\textbf{4-Order} } & R@20& 0.1068 & 0.1709& 0.0603& 0.1103 & 0.1083\\
                                         & N@20& 0.0884 & 0.2956& 0.0481& 0.1131 & 0.0679\\
  \bottomrule
\end{tabular}
}
\end{table}

 \begin{figure*}[h]
\begin{minipage}{\linewidth}
\center{
\includegraphics[width=.92\linewidth]{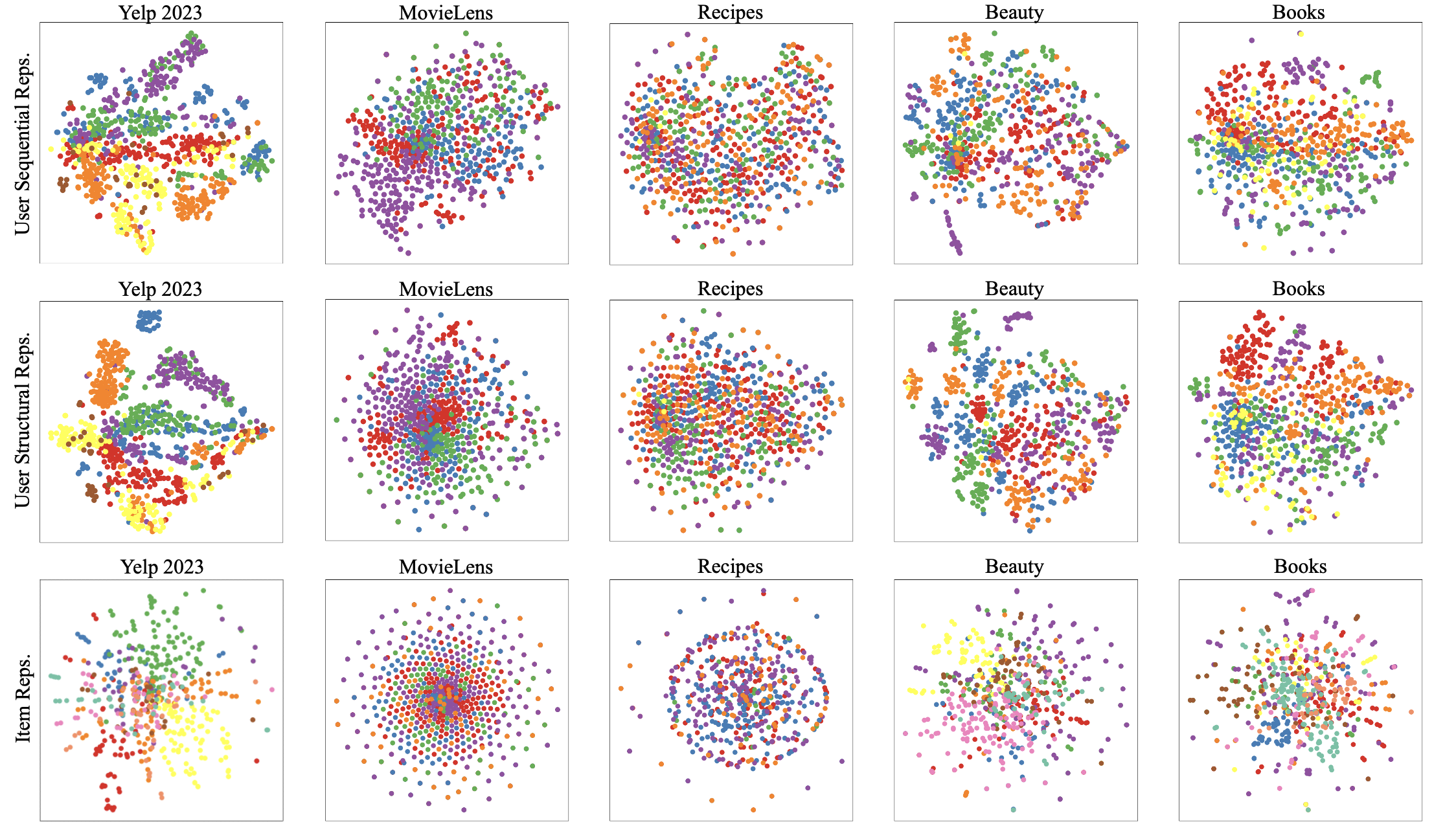}
  \caption{Visualizations of users’ sequential and structure representations (Reps.) and item representations.} 
  \label{fig:visualizations}
}
\end{minipage}
\end{figure*}

\subsection{In-Depth Analysis of NFARec}

To further understand the NFARec, we conducted further experiments. Due to the space limit, we excluded contents with similar observations, i.e., comparisons of sequential encoders on Books and Beauty datasets, and sensitivity of other hyperparameters.


\noindent
\textbf{Effect of Various Sequential Encoders.} We compared the encoder by NFARec with the Transformer, the neural Hawkes process, and the THP without masking. As illustrated in Table. \ref{tab:encoders}, we have the following findings: (1) the standalone Transformer as the encoder cannot attain the best performance, indicating that the sequential dependencies by Hawkes stochastic process cannot be ignored; (2) a major drawback of the neural Hawkes process is that it inherits the weaknesses of RNNs to capture long-term sequential dependencies; and (3) the encoder without masking will peek into the future and results in a representation degeneration. This indicates that sequentially predicting the sentiment polarity of the next interaction contributes to NFARec for recommendations.

\noindent
\textbf{Effect of Various Orders of Feedback.} It is interesting to note how high-order feedback impacts performance. The results in Table. \ref{tab:feedback_orders} provide the following findings: (1) the optimal performance is achieved with a 1-order feedback correlation on the Recipes and Books datasets, while a 2-order correlation works better on other datasets; and (2) excessive higher-order information diminishes the lower-order dependencies, affecting the performance. 

 \begin{figure}[t]
\begin{minipage}{\linewidth}
\center{
\includegraphics[width=\linewidth]{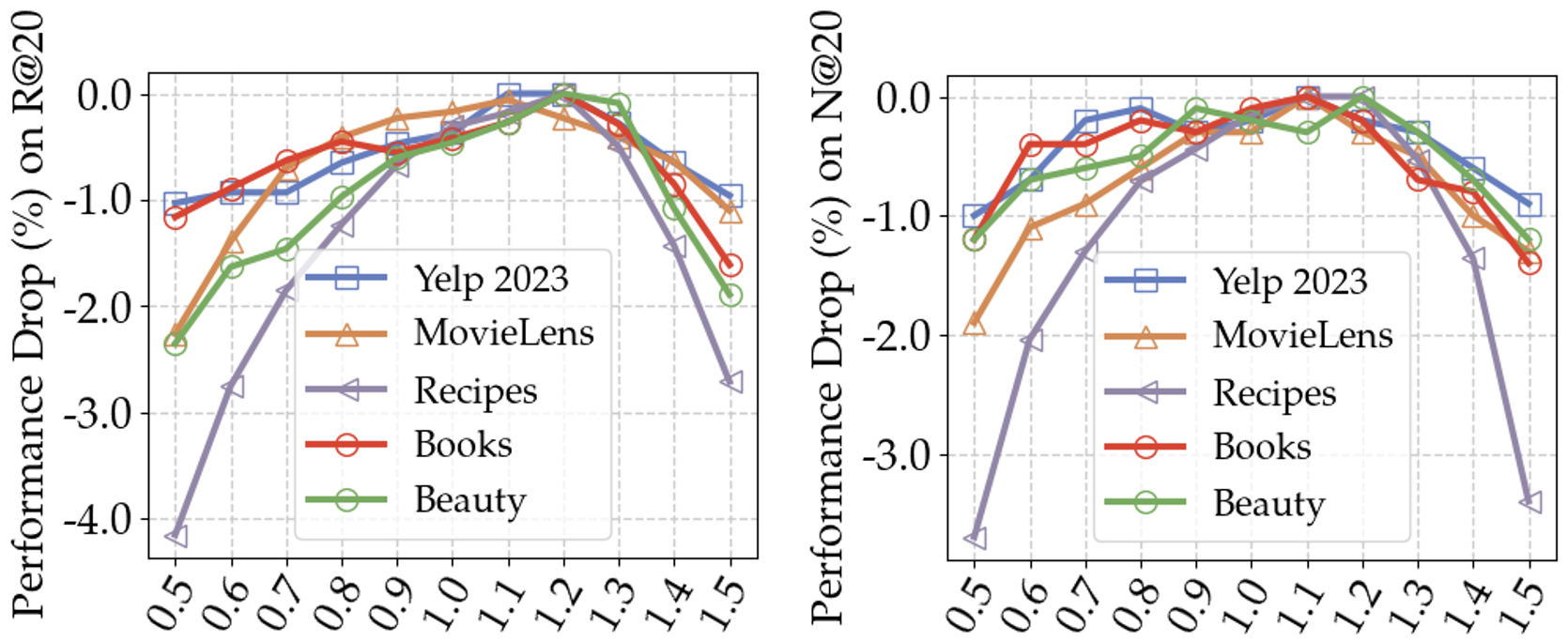}
  \caption{Effect of various $\delta$ on performance.} 
  \label{fig:delta}
}
\end{minipage}
\end{figure}

\noindent
\textbf{Computational Complexity.} 
The computational complexity of NFARec relies on three components, i.e., sequential representing learning ($\mathcal{O}$($|\bar{\mathcal{S}}|^2$), where $|\bar{\mathcal{S}}|$ denotes the average user interaction count), structure representation learning ($\mathcal{O}$($|\mathcal{I}|^2$)), and the decoder ($\mathcal{O}$($|\mathcal{I}|$)). As $|\mathcal{I}|^2$ $\gg$ $|\bar{\mathcal{S}}|^2$ + $|\mathcal{I}|$, the computational complexity of NFARec in a fixed setting is $\mathcal{O}$($|\mathcal{I}|^2$), indicating that considering the number of items is essential in a real deployment.

\noindent
\textbf{Parameter Sensitivity.} Fig. \ref{fig:delta} shows the effect of various $\delta$ on performance decline in terms of R@20 and N@20. We can see that (1) although there exists a slight distinction between Recall@20 and NDCG@20, the optimal value of $\delta$ is approximately 1.2 across all datasets. and (2) it is more sensitive on the Recipes dataset than others due to their different data distributions.



\begin{figure*}[h]
\center{
\begin{minipage}{\linewidth}
\includegraphics[width=\linewidth]{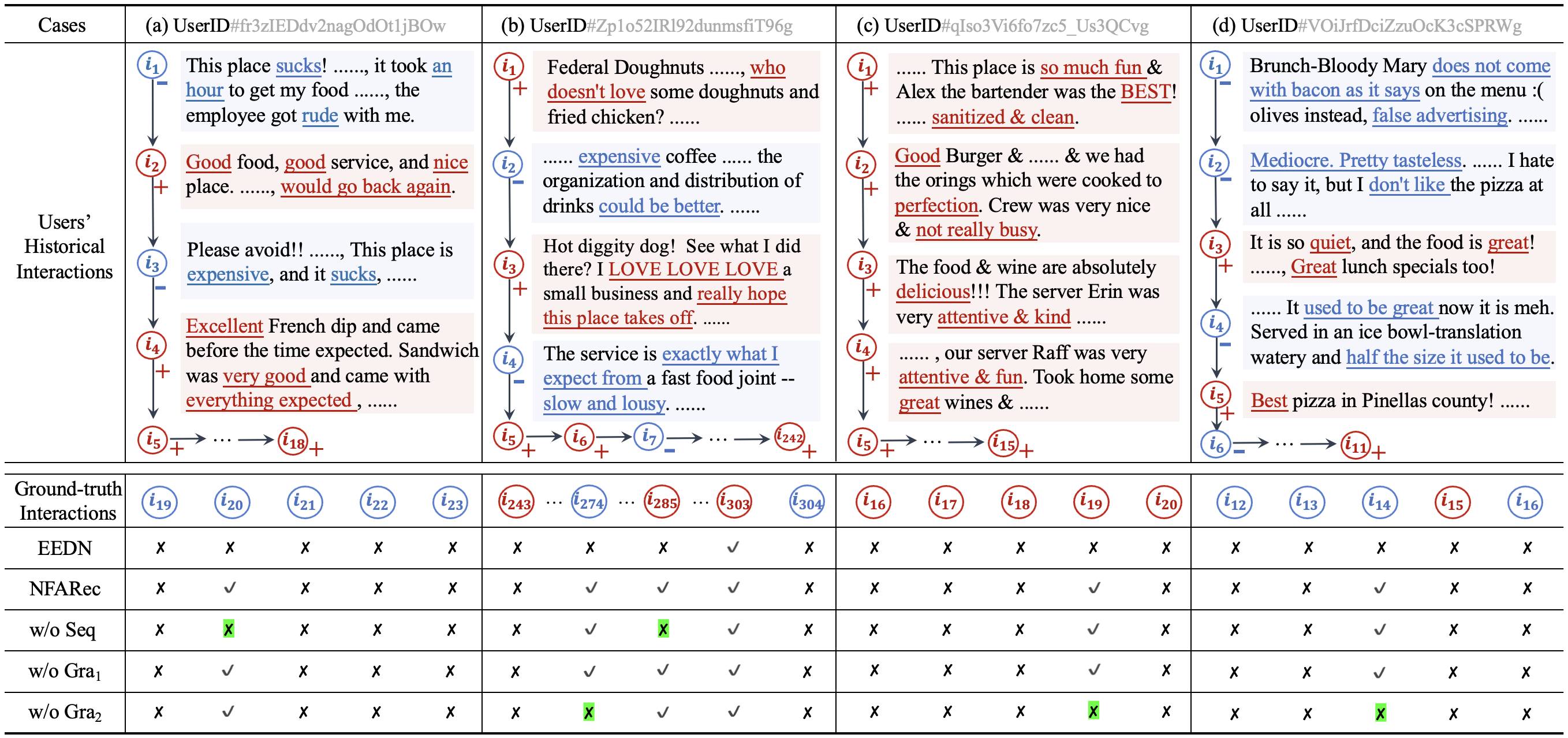}
  \caption{Case Study. {\footnotesize{\ding{52}}} and {\footnotesize{\ding{56}}} indicate that the model recommends correctly and incorrectly, respectively. \colorbox{green}{\footnotesize{\ding{56}}} denotes that the removed component is the key in each case. We do not show the case study on Gra$_1$, as it is not our main focus.} 
  \label{fig:case_study}
\end{minipage}
}
\end{figure*}

\subsection{Representation Visualization}
To examine the sequential and structural features, we visualize user and item representations in Fig. \ref{fig:visualizations}. Dots with the same color refer to users who interact with the same item or items that interact with the same user. 
 We have the following observations:

(1) In the majority of cases, NFARec effectively identifies the similarities between user and item representations, while the quality of visualization relies heavily on the datasets. Notably, the clustering results on the Yelp 2023, Beauty, and Books datasets outperform others by a significant margin.

(2) Both sequential and structural representations can reveal which users are similar but reflect different aspects. We can see that clusters of the same color in two representations exhibit distinct shapes and distributions on five datasets.


(3) The structural features suggest more cohesion within clusters, while sequential features exhibit greater dispersion. It indicates that the former reinforces the similarity and diversity among users, whereas the latter, by learning more uniform representations, helps alleviate the issue of inactive users becoming isolated nodes.


(4) Interestingly, the dense stacking of users and items on MovieLens is due to the large number of users engaging with the same items, averaging 165.6 shown in Table \ref{tab:statistics}. In contrast, the undesired visualization on the Recipes dataset results from a more challenging data distribution for recommendations. This is evident in the lower Recall and NDCG results in Table \ref{tab:comparison} compared to other datasets.

\subsection{Case Study}
We collected several apparent cases where NFARec achieves success, whereas baseline models fall short. We chose the Yelp 2023 dataset due to its explainable reviews. 


\noindent
\textbf{Cases for Sequential Representations.}
 We empirically found two major cases where the sequential representations are effective. 

 \noindent
 \textbf{Case 1.} The majority of the ground-truth candidates are with negative feedback. As Case (a) shown in Fig. \ref{fig:case_study}, the user has visited 18 places, and the ground-truth candidates from $i_{19}$ to $i_{23}$ were rated negatively by the user. In this scenario, NFARec successfully recommended item $i_{20}$, while EEDN incorrectly recommended high-rated items instead of the items that users will interact with. 
 
 \noindent
 \textbf{Case 2.} A user has a long historical visit sequence. In Case (b), the user has visited 242 places, and NFARec successfully recommended $i_{285}$. This is because the THP encoder, attributed to the Transformer, is crucial for capturing sequential correlations among nodes in long sequences to reflect their preferences. 

\noindent
\textbf{Cases for Structural Representations.}
 There are three typical cases to demonstrate the effectiveness of the feedback-aware structural representations.
 
 \noindent
 \textbf{Case 1.} A user exclusively visits places with high scores. In Case (c), the user rates all visited and future-visited places with positive scores, indicating that the user has less curiosity about unknown things. NFARec conveys information through an optimal feedback-aware path during HGCs and reduces message-passing between conflicting nodes, which is particularly beneficial for the user who solely entertains places with positive feedback.
 
 \noindent
 \textbf{Case 2.} 
A user actively explores new things. In Case (d), the majority of the user's experiences were with negative feedback, indicating that the user prefers to explore new places. For this user, the feedback-aware HGC operators in NFARec improve learning correlations between negative interactions, resulting in recommendations that better reflect real-life situations, even if the recommended items may be perceived negatively by the user.

 \noindent
 \textbf{Case 3.} 
A user is involved in an intricate interactive network. 
As depicted in Case (b) in Fig. \ref{fig:case_study}, the user has visited 242 places, signifying a substantial interactive network. The NFARec correctly recommends $i_{274}$ to the user based on structural representations. This indicates that NFARec can discriminate the user's latent preferences from the complex interactive graph.

\section{Conclusion}


In this paper, we proposed an NFARec that maximizes the utilization of negative feedback in sequential and structural patterns for recommendations. For sequential patterns, NFARec predicts the feedback sentiment of the next interaction based on the THP. This prompts the NFARec to identify items that the user has interacted with but may not prefer. For structural patterns, NFARec adopts a two-phase HGC approach to guide HGC operators to efficiently
exchange messages along feedback-aware paths. Extensive experiments on five public datasets show that NFARec outperforms SOTA methods. 
In the future, we intend to (1) explore the impact of implicit negative feedback to improve performance, such as skipping content, and (2) incorporate large language models to make more interpretable recommendations.

\section*{Acknowledgements}
We would like to thank anonymous reviewers for their thorough comments. This work is supported by the JKA Foundation (No.2024M-557) and China Scholarship Council (No.202208330093).

\newpage


\normalem
\bibliographystyle{ACM-Reference-Format}
\bibliography{sample-base}

\appendix

\end{document}